\documentclass[a4paper,12pt,tightenlines,notitlepage,nofootinbib,superscriptaddress]{revtex4-1}

\usepackage{amsmath,amssymb,setspace, soul}
\usepackage{graphicx,color}
\usepackage{geometry,layout}
\usepackage{newtxtext,newtxmath}
\usepackage[utf8x]{inputenc}
\usepackage[svgnames]{xcolor}
\usepackage[normalem]{ulem}
\usepackage[pdftex,unicode,colorlinks,citecolor=blue,linkcolor=blue,bookmarks=true]{hyperref}

\newcommand{\eg}{{\it e.g.\,}}

\newcommand{\Tr}{\mathop{\rm Tr}\nolimits}
\newcommand{\intinfty}{\displaystyle\int_{-\infty}^{\infty}\!}

\newcommand{\bra}[1]{\langle#1|}
\newcommand{\ket}[1]{|#1\rangle}

\begin{document}

\title{Quantum tomography enhanced through parametric amplification}

\author{E.~Knyazev}
\email{ev.knyazev@physics.msu.ru}
\affiliation{Faculty of Physics, M.V. Lomonosov Moscow State University, 119991 Moscow, Russia}

\author{K.~Yu.~Spasibko}
\affiliation{Max-Planck-Institute for the Science of Light, Staudtstrasse 2, 91058 Erlangen, Germany}
\affiliation{Friedrich-Alexander-Universit\"at Erlangen-N\"urnberg, Staudtstrasse 7/B2, 91058 Erlangen, Germany}

\author{M.~V.~Chekhova}
\affiliation{Faculty of Physics, M.V. Lomonosov Moscow State University, 119991 Moscow, Russia}
\affiliation{Max-Planck-Institute for the Science of Light, Staudtstrasse 2, 91058 Erlangen, Germany}
\affiliation{Friedrich-Alexander-Universit\"at Erlangen-N\"urnberg, Staudtstrasse 7/B2, 91058 Erlangen, Germany}

\author{F.~Ya.~Khalili}
\email{khalili@phys.msu.ru}
\affiliation{Faculty of Physics, M.V. Lomonosov Moscow State University, 119991 Moscow, Russia}
\affiliation{Russian Quantum Center, Skolkovo 143025, Russia}

\begin{abstract}

Quantum tomography is the standard method of reconstructing the Wigner function of quantum states of light by means of balanced homodyne detection. The reconstruction quality strongly depends on the photodetectors quantum efficiency and other losses in the measurement setup. In this article we analyse in detail a protocol of enhanced quantum tomography, proposed by Leonhardt and Paul in 1994 \cite{Leonhardt_PRL_72_4086_1994}, which allows one to reduce the degrading effect of detection losses. It is  based on phase sensitive parametric amplification, with the phase of the amplified quadrature being scanned synchronously with the local oscillator phase.  Although with sufficiently strong amplification the protocol enables overcoming any detection inefficiency, it was so far not implemented in experiment, probably due to the losses in the amplifier. Here we discuss a possible proof-of-principle experiment with a traveling-wave parametric amplifier. We show that with the state-of-the art optical elements, the protocol enables high-fidelity tomographic reconstruction of bright nonclassical states of light. We consider two examples: bright squeezed vacuum and squeezed single-photon state, with the latter being a non-Gaussian state and both strongly affected by the losses.

\end{abstract}

\maketitle

\section{Introduction}

During the last couple of decades, quantum optics experienced outstanding progress in the generation and application of non-classical light, which is considered as a necessary tool for many applications ranging from quantum information processing \cite{Gisin_RMP_74_145_2002, Kok_RMP_79_135_2007, Hammerer_RMP_82_1041_2010, Sangouard_RMP_83_33_2011, Andersen_NPhys_11_713_2015, Andersen_NPhys_11_713_2015} to high-precision interferometry \cite{Caves1981, Yurke_PRA_A_33_4033_1986, Nature_2011, Chekhova2016} and quantum optomechanics \cite{12a1Dakh, Aspelmeyer_RMP_86_1391_2014, 16a1DaKh},  including the preparation of mechanical objects in non-Gaussian quantum states \cite{Zhang_PRA_68_013808_2003, Mancini_PRL_90_137901_2003, 10a1KhDaMiMuYaCh, Romero-Isart_NJP_12_033015_2010}.

The standard method to characterize and verify the generated optical quantum states is quantum tomography \cite{Vogel_PRA_40_2847_1989, Smithey_PRL_70_1244_1993, Lvovsky_RMP_81_299_2009}, which allows one to reconstruct the Wigner function \cite{Wigner_PR_40_749_1932, Schleich2001} of a quantum state through balanced homodyne detection. The Wigner function possesses an important feature of a probability distribution: it yields marginal distributions of single variables. At the same time, unlike a ``true'' classical probability distribution, the Wigner function can take negative values, revealing the non-classical features of the corresponding states.

\begin{figure}
  \includegraphics[width=0.34\textwidth]{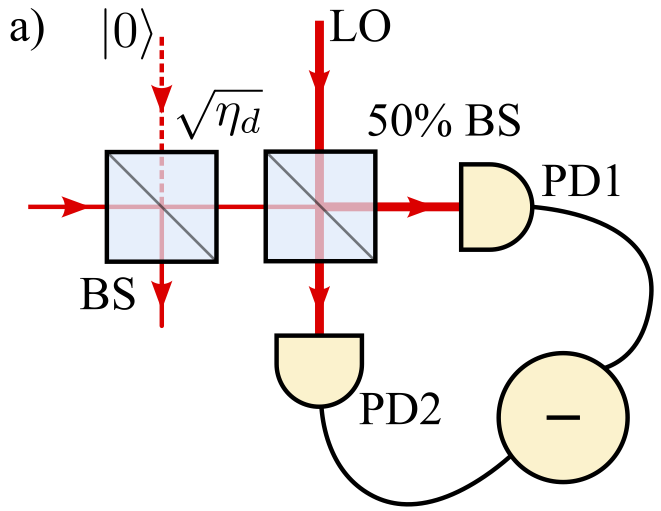}\qquad
  \includegraphics[width=0.48\textwidth]{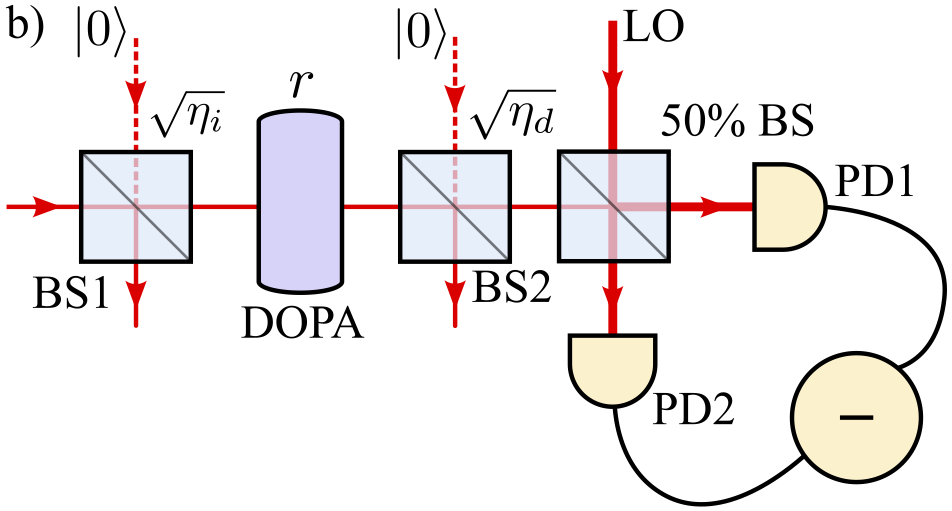}
  \caption{(a) Scheme of a balanced homodyne detector; (b) scheme of a parametric amplification enhanced homodyne detector. The detectors quantum inefficiencies and other optical losses are modelled by  beamsplitters BS, BS$1$, and BS$2$. The pump of the DOPA crystal is not depicted.}
  \label{fig:BHD}
\end{figure}

The idea of the homodyne tomography is shown in Fig.\,\ref{fig:BHD}\,a. Here the signal field is combined on a $50/50\%$ beamsplitter with a much stronger local oscillator field, forming the balanced homodyne detection scheme. For any given phase $\theta$ of the local oscillator, it measures the quadrature
\begin{equation}\label{q_theta}
  \hat{q}_\theta = \hat{q}\cos\theta + \hat{p}\sin\theta
\end{equation}
of the signal field defined by this phase ($q$ and $p$ are the dimensionless generalized position and momentum of the light mode). A set of probability distributions $w_\theta(q_\theta)$ for different quadratures  $q_{\theta}$ allows one to calculate the Wigner function using the inverse Radon transform.

The quality of the Wigner function reconstruction strongly depends on the optical losses in the  homodyne detection setup and, in particular, on the photodetectors quantum efficiency. The optical losses, by mixing the explored quantum state with the vacuum optical field, lead to the Gaussian blurring of the Wigner function, washing out its subtle details [see Eq.\,\eqref{tilde_W} below]. Bright nonclassical states of light are affected most strongly by this blurring. Note that although the best state-of-art homodyne detection setups reach $98.5\%$ detection efficiency \cite{Vahlbruch_PRL_117_110801_2016}, far better efficiencies could be needed to reconstruct the Wigner function of a bright non-Gaussian state, such as a strongly squeezed single photon state. Moreover, due to various technical reasons, like a non-perfect mode-matching, the overall detection efficiency in practical applications can be still as low as $\eta_d \sim0.5$ \cite{Nature_2011}, which is sufficient to completely remove the non-classical negative-valued area of the Wigner function of a single-photon state, even without squeezing.

A similar problem exists in high-precision optical interferometric phase measurements. In 1981 Caves proposed to use an anti-squeezer (a degenerate optical parametric amplifier, DOPA) at the output of a squeezed light fed interferometer in order to suppress the influence of the optical losses in the output path \cite{Caves1981}. Later this idea was further developed in several papers \cite{Yurke_PRA_A_33_4033_1986, Marino_PRA_86_023844_2012, 17a1MaKhCh} and demonstrated experimentally \cite{Jing_APL_99_011110_2011, Kong_APL_102_011130_2013, Hudelist_NComms_5_3049_2014, 17a1MaLoKhCh}.

 In 1994, Leonhardt and Paul proposed to use the same pre-amplification principle in homodyne optical tomography \cite{Leonhardt_PRL_72_4086_1994}, see Fig.\,\ref{fig:BHD}\,b. Here, a phase-sensitive amplifier (a DOPA) is used to amplify the measured quadrature $\hat{q}_\theta$. Evidently, to achieve this goal, the amplification phase has to be synchronized with the local oscillator phase.

However, this approach requires non-linear optical element(s) which are typically much more lossy than linear ones. A serious problem is that at least part of the vacuum noise caused by these losses is amplified together with the incident light and therefore influences the performance much more strongly than the ordinary inefficiency of detectors.

Here, we analyze a possible proof-of-principle {\it enhanced homodyne tomography} experiment \cite{Leonhardt_PRL_72_4086_1994}, taking into account losses in the DOPA. As for the DOPA, we consider a traveling-wave parametric amplifier based on a $\chi^{(2)}$ nonlinear crystal pumped with strong pulses. We show that using the state-of-the-art optical elements, it is possible to significantly improve the quality of the Wigner function reconstruction.

This article is organized as follows. In the next section we briefly review the principle of the quantum tomography and discuss the role of the optical losses.
In Sec.\,\ref{sec:EnhTom} we analyze the enhanced homodyne tomography scheme of \cite{Leonhardt_PRL_72_4086_1994}, taking into account the losses in the amplifier. Then in Sec.\,\ref{sec:exper} we discuss a possible proof-of-principle experiment and consider examples of bright squeezed vacuum and squeezed single-photon states.

\section{Homodyne tomography}\label{sec:BasicTheory}

\subsection{Lossless case}\label{subseq:lossless}

In order to provide the reference point for our consideration below, we start with an ideal lossless case. Following the seminal paper \cite{Vogel_PRA_40_2847_1989}, we use the convenient and mathematically transparent approach, based on the characteristic functions.

The characteristic function of the quadrature \eqref{q_theta} is defined as
\begin{equation}\label{C_theta_rho}
C_\theta(\xi) = \Tr\left(\hat{\rho}e^{i\xi\hat{q}_\theta}\right) = \text{Tr} \left( \hat{\rho} e^{i\xi (\hat{{\rm a}}e^{-i\theta} + \hat{{\rm a}}^\dagger e^{i\theta})/\sqrt{2}} \right),
\end{equation}
where $\hat{\rho}$ is the density operator of the quantum state of an optical mode and $\hat{{\rm a}}$ is the annihilation operator of this mode. One can show that it is equal to the Fourier transform of the probability distribution $w_\theta(q_\theta)$ of $\hat{q}_\theta$:
\begin{equation}\label{w2C_theta}
  C_\theta(\xi) = \intinfty w_\theta(q_\theta)e^{i\xi q_\theta}\,dq_\theta \,.
\end{equation}

On the other hand, the Wigner function $W(q,p)$ can be defined as the inverse Fourier transform of the symmetrized joint characteristic function $C(z)$ for $q$ and $p$ \cite{Vogel_PRA_40_2847_1989}:
\begin{equation}\label{C2W}
  W(q,p) = \frac{1}{(2\pi)^2}\intinfty C(z)e^{-i(z'q+z''p)}dz'dz'' \,,
\end{equation}
where
\begin{equation}\label{rho2C}
  C(z) = \Tr\left(\hat{\rho} e^{i(z'\hat{x} + iz''\hat{p})}\right)
  = \Tr\left(
        \hat{\rho}
          e^{i (z^\prime\hat{{\rm a}} + z\hat{{\rm a}}^\dagger)/\sqrt{2}}
      \right),
\end{equation}
and $z = z' + iz''$. It follows from Eqs.\,(\ref{C_theta_rho}, \ref{rho2C}) that
\begin{equation}\label{C_theta2C}
  C(\xi e^{i\theta}) = C_\theta(\xi) \,.
\end{equation}
The chain of equations \eqref{w2C_theta}$\to$\eqref{C_theta2C}$\to$\eqref{C2W} constitutes in essence the inverse Radon transform.

\subsection{Detection losses} \label{subsec:DetLoss}

We model the quantum inefficiency of the homodyne detector scheme by means of an imaginary beamsplitter BS, see Fig.\,\ref{fig:BHD}\,a, which mixes the input field with the vacuum field having the annihilation operator $\hat{{\rm z}}_d$ \cite{Leonhardt_PRA_48_4598_1993}. The density operator of the resulting damped quantum state can be represented as (we denote the damped state with the prime throughout the article) \begin{equation}\label{tilde_rho}
  \hat{\rho}^\prime = \Tr_{\rm z}\left(
    \hat{U}_d \hat{\rho}\otimes\ket{0}\bra{0}
     \hat{U}^\dagger_d 
  \right) ,
\end{equation}
where $\Tr_{\rm z}$ is the partial trace taken over the vacuum field subspace and the unitary operator $\hat{U}_d$ describes the action of the BS. In particular, the annihilation operator of an incident field is transformed as
\begin{equation}\label{calU_d}
  \hat{U}_d^\dagger\hat{{\rm a}}\hat{U}_d
  = \sqrt{\eta_d}\,\hat{{\rm a}} + \sqrt{1-\eta_d}\,\hat{{\rm z}}_d \,,
\end{equation}
 where $\eta_d$ is the power transmissivity of the BS. The corresponding characteristic function for the damped quadrature $\hat{q}_\theta$ is
\begin{equation}\label{tilde_theta_C}
  C^\prime_\theta(\xi)
  = \Tr\left(\hat{\rho}^\prime e^{i\xi\hat{q}_\theta}\right)
  = C_\theta(\sqrt{\eta_d}\,\xi) e^{-(1-\eta_d)\xi^2/4} .
\end{equation}
Direct application of relation \eqref{C_theta2C} to this characteristic function yields the Wigner function of the damped quantum state, see \eg Eq.\,(18) of \cite{Lvovsky_RMP_81_299_2009}.

The structure of \eqref{tilde_theta_C} reflects the interplay of the attenuation and the noise which takes place in any lossy dynamics. The scaling factor $\sqrt{\eta_d}$ describes the irreversible attenuation, while the exponential one --- the Gaussian blurring of the Wigner function due to the injected vacuum noise. This approach yields values of $q$ and $p$ shrunk by a factor of $\sqrt{\eta_d}$. In order to restore the initial quantum state with high fidelity, it is natural to rescale them back by using, instead of \eqref{C_theta2C}, the equation
\begin{equation}\label{C_theta2Closs}
 C^\prime(\sqrt{\eta_d}\,\xi e^{i\theta}) = C^\prime_\theta(\xi) \,,
\end{equation}
where
\begin{equation}\label{tilde_C}
  C^\prime(z) = C(z) e^{- \epsilon_d|z|^2 / 4}
\end{equation}
 is the {\it unbiased reconstruction} of the initial characteristic function $C$, and $\epsilon_d = (1-\eta_d) / \eta_d$ is the normalized loss factor. The term ``unbiased'' means that $C^\prime$ gives the correct mean values of $q$, $p$. Note that ``unbiased'' reconstruction of the Wigner function does not yield the quantum state after the loss, but is merely a convenient practical procedure for estimating the Wigner function of the input quantum state.

From the experimental point of view the rescaling described by Eq.\eqref{C_theta2Closs} is tolerant to errors in the estimation of detector efficiency $\eta_d$. These errors affect only the mean values of $q$, $p$, not the structure of the Wigner function. On the contrary, an attempt to undo the noise influence by multiplying $C^\prime_\theta$ in (\ref{tilde_theta_C}) by $e^{(1-\eta_d)\xi^2/4}$ requires very precise knowledge of $\eta_d$ (note that the multiplication factor increases exponentially with $\xi$). Therefore, this option is not considered as a practical one, see \eg \cite{Leonhardt_PRL_72_4086_1994}.

The inverse Fourier transform of \eqref{tilde_C} gives the corresponding {\it unbiased reconstruction} of the initial (before loss) Wigner function:
\begin{equation}\label{tilde_W}
  W^\prime(q,p) = \intinfty C^\prime(z)e^{-i(z'q+z''p)}\,\frac{dz'dz''}{(2\pi)^2}
  = \intinfty W(q',p')B(q-q',p-p')dq'dp' \,,
\end{equation}
where
\begin{equation} \label{eq:GaussBlur}
  B(q,p) = \frac{1}{\pi\epsilon_d} e^{-\left(q^2+p^2\right)/\epsilon_d}
\end{equation}
is the Gaussian blurring kernel.

\section{Enhanced homodyne tomography}\label{sec:EnhTom}

Now, following \cite{Leonhardt_PRL_72_4086_1994}, suppose that a degenerate parametric amplifier with a gain $r$ is added to our scheme, see  Fig.\,\ref{fig:BHD}\,b. In a real-world experiment one has to take into account the losses in the DOPA itself. In the general case, three kinds of losses have to be distinguished: (i) the input loss, whose noise is amplified by the DOPA to the same extent as the incident optical field; (ii) the loss inside the DOPA, whose noise is partly amplified; and (iii) the output loss, whose noise is not amplified. In the case of a single-pass DOPA, they correspond to, respectively: (i) absorption and reflection in the input anti-reflective coating of the nonlinear crystal; (ii) absorption in the crystal bulk; and (iii) absorption and reflection of the output coating. It is evident that the latter can be included into the detector inefficiency, therefore, below we will not consider it separately.

In Fig.\,\ref{fig:BHD}\,b, the DOPA input losses and the detector inefficiency (including the DOPA output losses) are modeled by beamsplitters BS1 and BS2 with the power transmissivities $\eta_i$ and $\eta_d$, correspondingly, located in front of and on the rear of the DOPA.

With an account for all these losses and the parametric amplification, Eq.\,\eqref{tilde_rho} takes a more sophisticated form,
\begin{equation}\label{tilde_rho_amp}
  \hat{\rho}^\prime = \Tr_{\rm z}\left[
   \hat{U}_d\hat{S}(r,\theta)\hat{U}_i
      \hat{\rho}\otimes\ket{0}\bra{0}
      \hat{U}_i^\dagger\hat{S}^\dagger(r,\theta)
      \hat{U}_d^\dagger
    \right] .
\end{equation}
Here $\hat{U}_i$ is the evolution operator of damping defined similar  to \eqref{calU_d},
\begin{equation}
 \hat{U}_i^\dagger\hat{{\rm a}}\hat{U}_i
  = \sqrt{\eta_i}\,\hat{{\rm a}} + \sqrt{1-\eta_i}\,\hat{{\rm z}}_i \, ,
\end{equation}
and $\hat{S}$ is the squeezing operator which also takes into account the absorption inside the DOPA:
\begin{equation}
 \hat{S}^\dagger(r,\theta)\hat{q}_\theta\hat{S}(r,\theta)
  = ( \hat{q}_\theta  +  \hat{q}_a ) e^r \,,
\end{equation}
where $ \hat{q}_a$ is the quadrature of the introduced noise translated to the DOPA input, and $r$ is the effective squeezing defined in \eqref{eq:Reff}. The explicit form of $ \hat{q}_a$ for the case of a single-pass DOPA is calculated in App.\,\ref{app:singlepass}, with the variance $\sigma_a^2$ given by
\begin{equation} \label{appeq:BulkNoiseVar}
	  \sigma_a^2 = \frac{k d}{4 r} \left( 1- e^{-2 r} \right) \, ,
\end{equation}
where $k$ and $d$ are correspondingly absorption coefficient and length of the crystal. The above equations give the following characteristic function of quadrature $q_\theta$:
\begin{equation}\label{tilde_theta_C_app}
  C^\prime_\theta(\xi)
  = C_\theta(\sqrt{\eta_i\eta_d}\,\xi  e^r)\exp\left[-\frac{\xi^2}{2}\left(
        \frac{(1-\eta_i)\eta_d}{2}\,  e^{2r} + \sigma_a^2  \eta_d e^r + \frac{1-\eta_d}{2}
      \right)\right] \,.
\end{equation}

In order to obtain the Wigner function of the input state, we use the same rescaling (``unbiased'') approach as in Sec.\,\ref{subsec:DetLoss}. It is especially justified in the case of amplification due to the large factor $e^{2r}$. Namely, we assume that
\begin{equation}\label{C_theta2Capp}
  C^\prime(\sqrt{\eta_i\eta_d}\,  e^r\xi e^{i\theta}) = C^\prime_\theta(\xi) \,.
\end{equation}
In this case,
\begin{equation}\label{tilde_C_amp}
  C^\prime(z) = C(z)e^{-\epsilon_r|z|^2/4} \,,
\end{equation}
where
\begin{equation}\label{eps_r}
  \epsilon_r = \epsilon_i + \frac{2\sigma_a^2 + \epsilon_d e^{-2r}}{\eta_i}
\end{equation}
is the new effective loss factor and $\epsilon_i = (1-\eta_i) / \eta_i$ is its component due to the input losses. Note, that the noise stemming from the detector inefficiency is suppressed exponentially in $r$, that is linearly in the amplification factor. The noise created by the crystal absorption is also suppressed, but only logarithmically in the amplification factor, see Eq.\,\eqref{appeq:BulkNoiseVar}. Nevertheless, for relatively high $r$, the main noise influence comes from the input losses, which can be extremely small, see Sec.\,\ref{sec:exper}.

As in the case of \eqref{C_theta2Closs}, the experimental ``unbiased'' reconstruction requires approximate knowledge of $\sqrt{\eta_i\eta_d}\, e^r$ factor. The main part $\sqrt{\eta_d}\, e^r$ can be obtained from the calibration measurement with the vacuum field as an input state, and the input losses $\eta_i$ can be estimated by means of a separate measurement, e.g. by measuring the reflectivity of the crystal surface.

Finally, the inverse Fourier transform of \eqref{tilde_C_amp} gives the same result \eqref{tilde_W} for the Wigner function reconstruction, but with the blurring kernel depending on the loss factor $\epsilon_r$:
\begin{equation}
 B(q,p) = \frac{1}{\pi\epsilon_r} e^{-\left(q^2+p^2\right)/\epsilon_r} .
\end{equation}

\section{Performance and estimates} \label{sec:exper}

It follows from the above analysis that the pre-amplification will dramatically improve the homodyne tomography of any quantum state whose features are strongly affected by losses. In particular, Eq.\,\eqref{eps_r} means that in principle, any detection inefficiency can be compensated for by the sufficiently strong amplification (anti-squeezing). However, due to the structure of (\ref{eps_r}), only large values of amplification could compensate for the losses in the amplifier itself, therefore very accurate experimental planning is required in order to obtain a high fidelity of reconstruction.

Throughout this section, we consider a BBO crystal of length $d=1\,\text{mm}$ for the phase-sensitive pre-amplification, with the value of bulk absorption of $k=0.1\,\text{m}^{-1}$ \cite{dmitriev2013handbook}. We also consider the commercially available anti-reflective coating with the reflectivity $\eta_i=99.99\%$ at $800 \, \text{nm}$ \cite{newlightphotonics_bbo}. All figures and estimations are given for these parameters of crystal reflection and absorption.

\subsection{Bright squeezed vacuum}

\begin{figure}
  \includegraphics[width=1\textwidth]{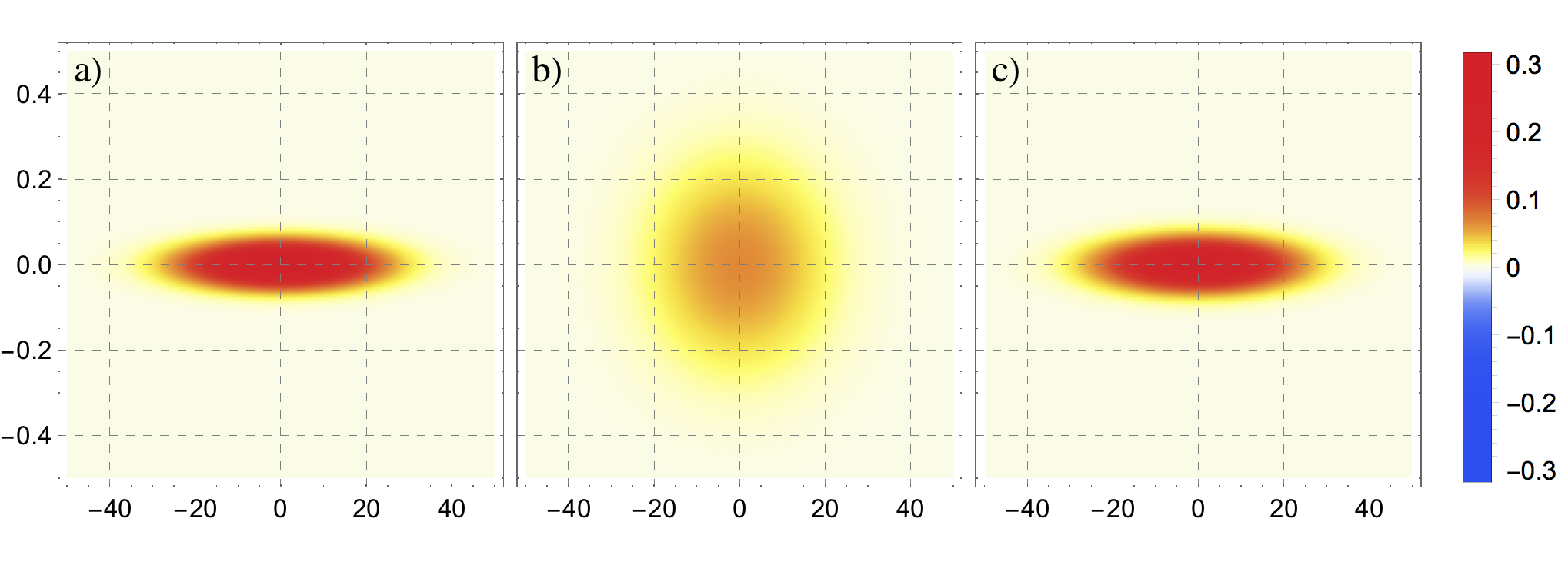}
  \caption{The Wigner function of bright squeezed vacuum (BSV): a) BSV state \eqref{W_BSV} with the preparation gain $r_1=3$ (26\,db of squeezing); b) the Wigner function reconstruction \eqref{tilde_W} with $\eta_d=95\%$ detection efficiency; c) reconstruction with  $\eta_d=95\%$ quantum efficiency after 20\,db of amplification ($r_2=2.3$). Graphs are produced for imperfect BBO non-linear crystal. Note that for better visualisation, the axes are scaled disproportionally.}
  \label{fig:BrightSqzVac}
\end{figure}

The simplest case is bright squeezed vacuum, which can be generated at the output of an unseeded strongly pumped DOPA. This is a Gaussian state with the  Wigner function
\begin{equation}\label{W_BSV}
  W_{\rm BSV}(q,p) = \frac{1}{\pi} e^{-q^2s^{-1} - p^2s} ,
\end{equation}
with $s=e^{2r_1}$, $r_1$ being the parametric gain, which also determines the mean number of photons $N=\sinh^2r_1$.

Consider a state with the mean number of photons $N\approx100$, which corresponds to a strong squeezing of 26\,dB, that is $r_1\approx3$ (note that much stronger squeezing is achievable by pumping a BBO crystal with picosecond pump pulses of about $10^2\,\mu{\rm J}$ energy \cite{Iskhakov_OL_37_1919_2012}). The corresponding Wigner function is shown in Fig.~\ref{fig:BrightSqzVac}\,a. However, this impressive degree of squeezing is impossible to observe in practice: a detection loss $1-\eta_d$ exceeding $e^{-2r_1}\approx0.0025$ will almost completely destroy the purity of the state. Homodyne detection after such a loss will retrieve not the squeezed quadrature uncertainty, but mainly the amount of loss, see Fig.~\ref{fig:BrightSqzVac}\,b, where the  rescaled reconstruction of $W_\text{BSV}$ \eqref{tilde_W} is plotted for the case of $1-\eta_d=0.05$.

At the same time, if the quadrature under measurement is amplified before the homodyne detection, by sending the state to another DOPA with a sufficiently large parametric gain $r_2$, both the initially squeezed and the initially anti-squeezed quadratures will become anti-squeezed, the Wigner function distribution will be simply rescaled, and the measurement will correctly retrieve its aspect ratio $e^{2r_1}$ and hence the degree of squeezing.

Indeed, the reconstructed Wigner function after the amplification, with losses taken into account, is
\begin{equation}\label{tilde_W_BSV}
  W^\prime_{\rm BSV}(q,p)
  = \frac{\exp\left(-\dfrac{q^2}{\epsilon_r+s} - \dfrac{p^2}{\epsilon_r+s^{-1}}\right)}
      {\pi\sqrt{(\epsilon_r+s)(\epsilon_r+s^{-1})}} \,,
\end{equation}
 where  $\epsilon_r$ is given  by (\ref{eps_r}), (\ref{appeq:BulkNoiseVar}), and (\ref{eq:Reff}) with $r=r_2$. We see that if $\epsilon_r  \ll s^{-1}$, then the reconstruction \eqref{tilde_W_BSV} reduces to the initial Wigner function \eqref{W_BSV}.

In Fig.\,\ref{fig:BrightSqzVac}\,c, the reconstructed Wigner function \eqref{tilde_W_BSV} is plotted for the case of $r_2=2.3$ (20\,db). One can see that although only $1-\eta_d=5\%$ of detection loss completely change the initial shape of the distribution, the relative moderate  pre-amplification enables the reconstruction of the shape. Even though the reflection loss is irreversible, modern experimental techniques allow one to achieve very low amount of input loss, therefore enabling the efficient experimental implementation of the squeezing enhancement tomography protocol.

\begin{figure}
  \includegraphics[width=0.62\textwidth]{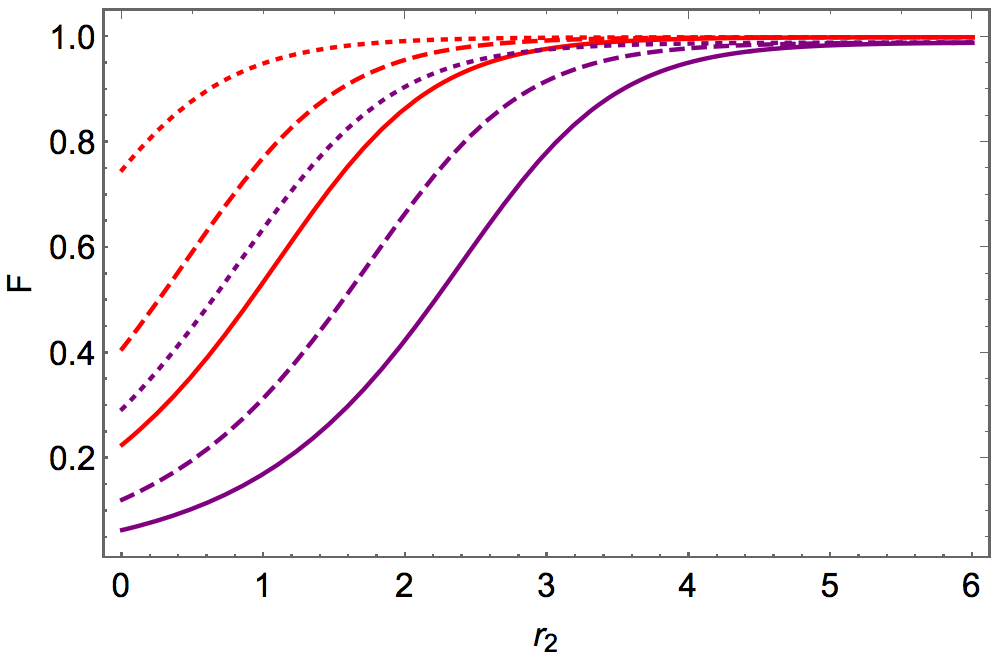}
  \caption{Fidelity (\ref{eq:Fdl}) of reconstructing a bright squeezed vacuum state. Red curve: $r_1=1.7$ for $\eta_d=45\%$ (solid), $\eta_d=75\%$ (dashed) and $\eta_d=95\%$ (dotted) detector efficiency. Purple curve: $r_1=3$ for $\eta_d=45\%$ (solid), $\eta_d=75\%$ (dashed) and $\eta_d=95\%$ (dotted) efficiency of photodetector. Note slightly smaller value of limiting fidelity for a stronger squeezed state: this is due to the irreversible input loss.}
  \label{fig:Fidelity}
\end{figure}

\begin{figure}
  \includegraphics[width=0.72\textwidth]{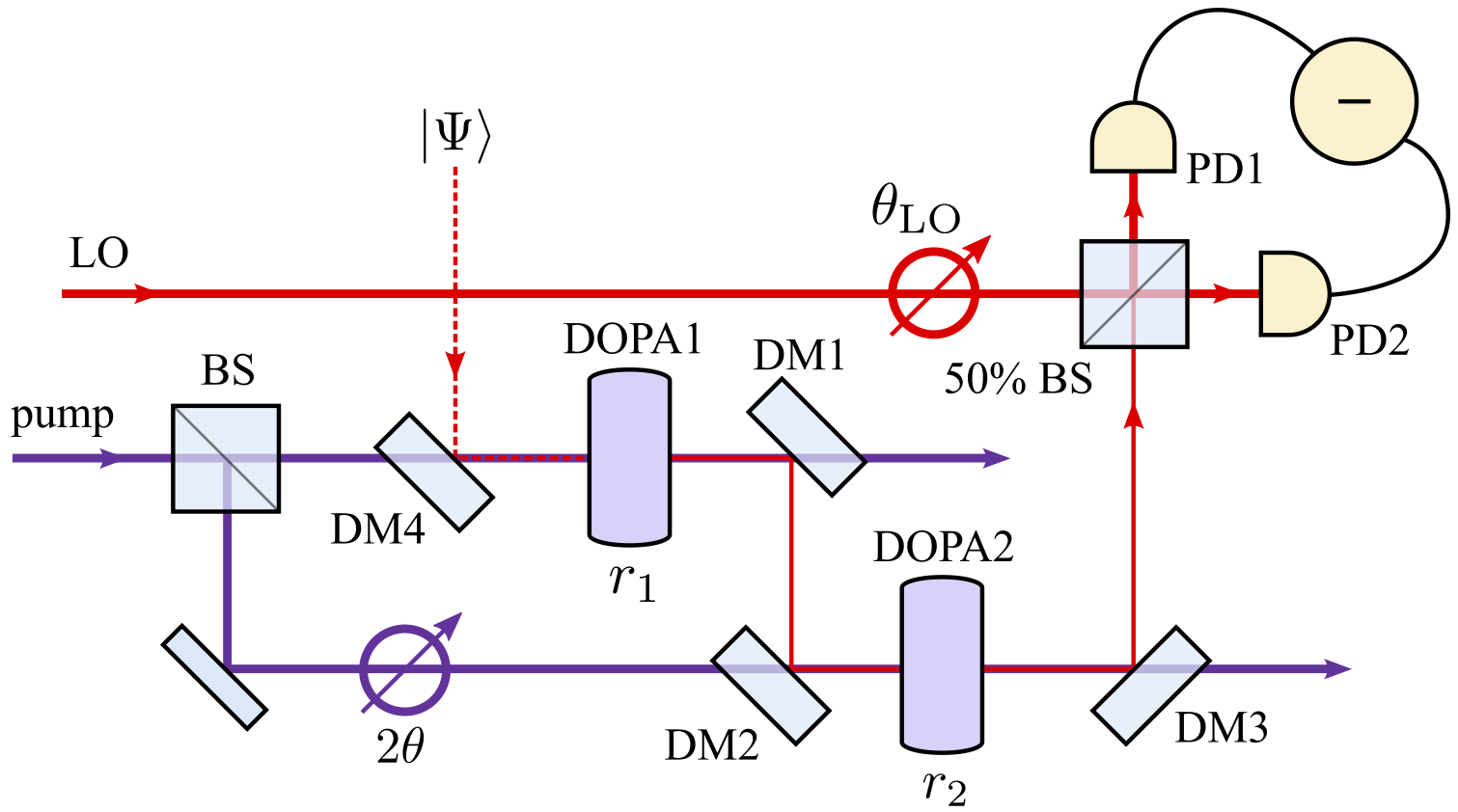}
  \caption{A possible experimental scheme for the enhanced tomography of bright squeezed vacuum or a squeezed non-Gaussian state. Bright squeezed vacuum is prepared in the phase-sensitive amplifier DOPA1 without seeding, whilst the squeezed non-Gaussian state can be obtained by seeding the DOPA1 by a non-Gaussian (for instance, single-photon) state $|\Psi\rangle$. Homodyne tomography is performed using the local oscillator LO whose phase $\theta_\text{LO}$ can be scanned, and a balanced beamsplitter 50\% BS. Before the homodyne detection, the quadrature under measurement is amplified by DOPA2, with the proper quadrature chosen by scanning one of the interferometer mirrors. DM1-DM4 are dichroic mirrors transmitting the pump and reflecting the down-converted radiation. }
  \label{fig:setup}
\end{figure}

A certain difficulty  while reconstructing a strongly squeezed state will arise due to the narrow range of phases for which squeezing can be observed. Within this range, whose width is given by the inverse aspect ratio of the Wigner function distribution, the phase should be scanned with a very high resolution: in this example, about $10^{-3}$ rad.

As a measure of reconstruction quality, we consider the fidelity, defined as the overlap of the corresponding Wigner functions. In the simple case of a squeezed vacuum state, the analytical expression for the fidelity can be obtained:
\begin{equation} \label{eq:Fdl}
F = \frac{2}{\sqrt{(\epsilon_r e^{2r_1}+2)(\epsilon_r e^{-2r_1}+2)}}.
\end{equation}
This parameter is plotted in Fig.\,\ref{fig:Fidelity} as a function of the phase-sensitive amplification gain $r_2$. One can notice that strong pre-amplification allows one to achieve a high, but limited value of fidelity, which is due to the irreversible reflection loss $\eta_i \neq 1$. Anyway, for the state-of-the-art anti-reflection coatings and high-gain parametric amplification, the satisfactory value of $F \approx 0.99$ is achievable.

Figure~\ref{fig:setup} shows a possible experimental setup. Both amplifiers, DOPA1 generating bright squeezed vacuum and  DOPA2 amplifying the quadrature under measurement, are parts of an SU(1,1) interferometer~\cite{17a1MaKhCh}, in which the pump power can be distributed unequally by the beamsplitter BS, and hence the parametric gain values $r_{1,2}$ could be different. The state produced by DOPA1 is sent  for amplification to DOPA2 through dichroic mirrors DM1 and DM2, while the amplification phase $\theta$ is varied in the pump beam synchronously with the phase $\theta_\text{LO}$ of the local oscillator LO used for homodyne detection.

\subsection{Squeezed single-photon state}

\begin{figure}
  \includegraphics[width=1\textwidth]{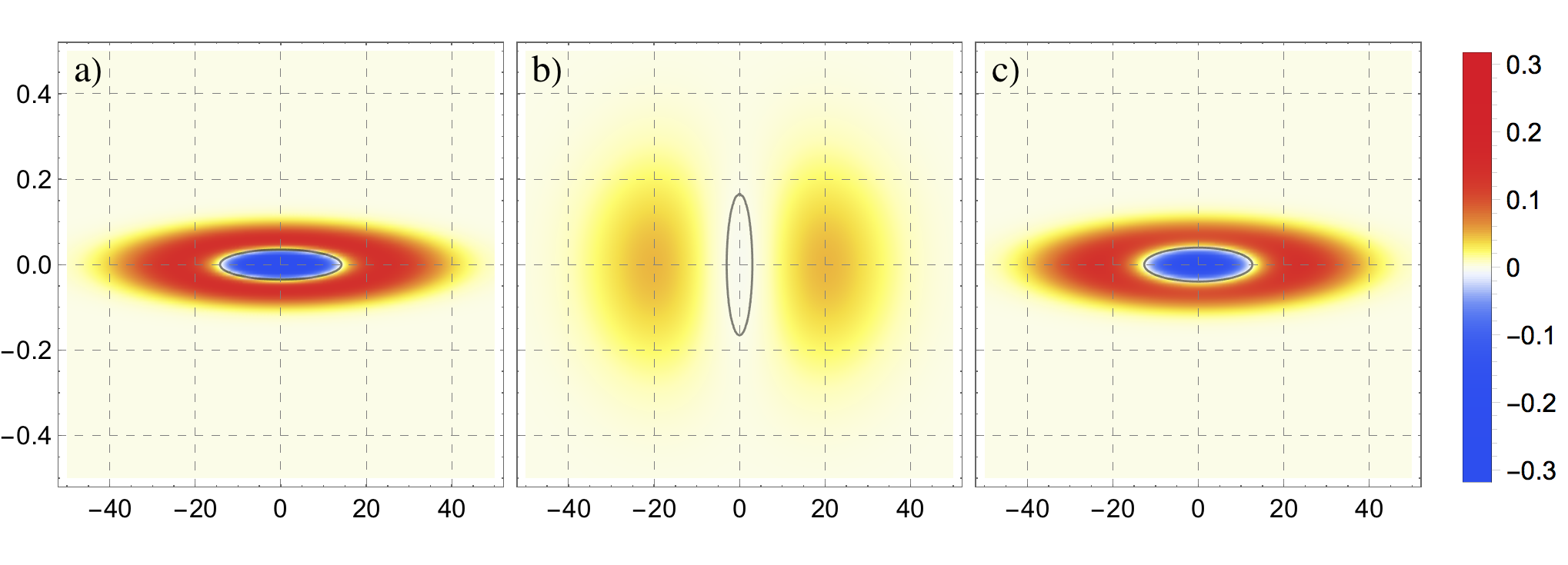}
  \caption{a) The Wigner function of a squeezed single-photon state \eqref{W_SSP} with $r_1=3$ (26\,db of squeezing); b) the Wigner function reconstruction \eqref{tilde_W} with $\eta_d=95\%$ detection efficiency; c) the reconstruction with $\eta_d=95\%$ quantum efficiency after 20\,db of pre-amplification ($r_2=2.3$). Black ellipses encircle the negative-valued area.  Graphs are produced for imperfect BBO non-linear crystal. Note that for better visualisation, the axes are scaled disproportionally.}
\label{fig:WignerSinglePhot}
\end{figure}

Even more dramatic is the effect of phase-sensitive amplification on the homodyne tomography of bright non-Gaussian states, for example, a squeezed single-photon state (SSP) $\hat{S}\ket{1}$,   see the review papers \cite{Lvovsky_RMP_81_299_2009, Andersen_NPhys_11_713_2015} and the references therein, as well as the recent works \cite{Miwa_PRL_113_013601_2014, Baune_PRA_95_061802_2017}. Its Wigner function has the form
\begin{equation}\label{W_SSP}
  W_{\rm SSP}(q,p) = \frac{2s^{-1} q^2 + 2s p^2 -1}{\pi}\,
    \exp\left(-q^2s^{-1}- p^2s\right),
\end{equation}
and it contains a negative-valued area stretched along one quadrature and squeezed along the other, see Fig.\,\ref{fig:WignerSinglePhot}\,a.

\begin{figure}
  \includegraphics[width=1\textwidth]{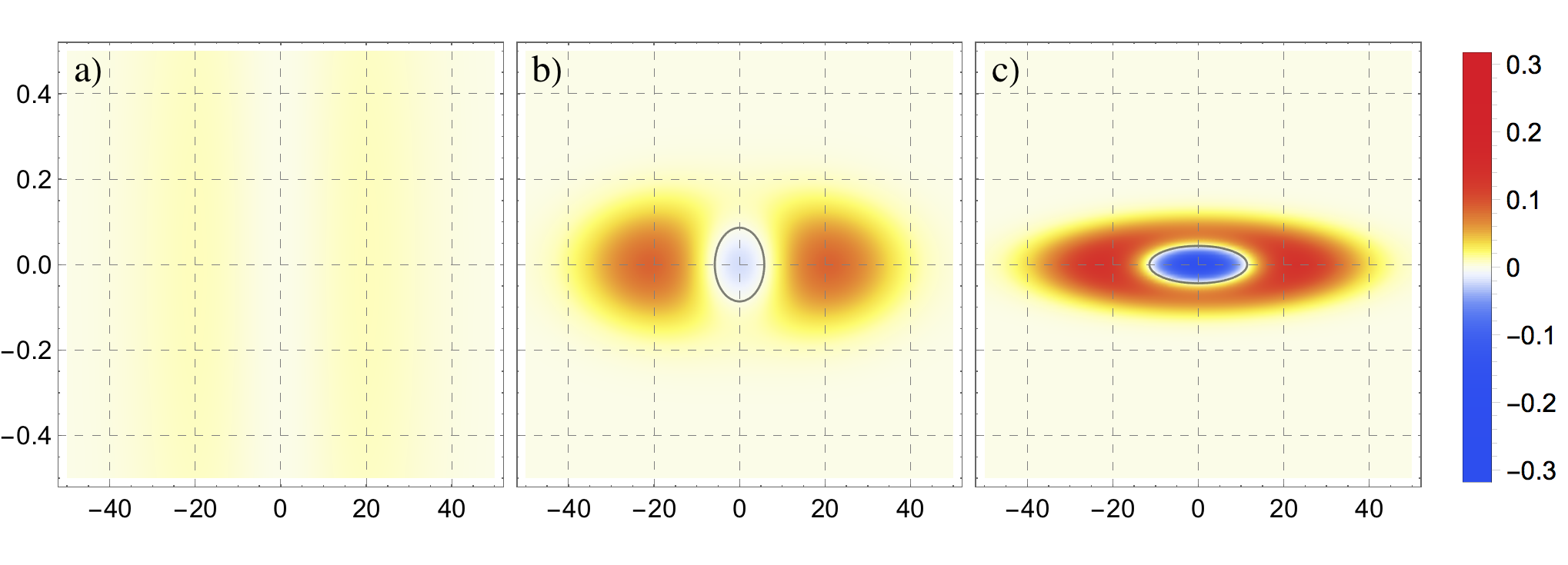}
  \caption{Reconstruction of the Wigner function of squeezed single-photon state with  $r_1=3$ (26\,db of squeezing) in the case of low detection efficiency. a) Reconstruction \eqref{tilde_W} with $45\%$ detection efficiency; b) reconstruction with 45\% detection efficiency after 20\,db of amplification ($r_2=2.3$); c) reconstruction with 45\% detection efficiency after 30\,db of amplification ($r_2=3.45$). Black ellipses encircle the negative-valued area. Graphs are produced for the case of an imperfect BBO non-linear crystal.  Note that for better visualisation, the axes are scaled disproportionally.}
\label{fig:WignerSinglePhot06}
\end{figure}

This state can be generated by seeding DOPA1 in Fig.~\ref{fig:setup} with single photons prepared, for instance, through the heralding procedure \cite{Lvovsky_PRL_87_050402_2001}.  We assume the same degree of the preparation squeezing as in the previous example: $r_1\approx3$ (26 db), which corresponds to the mean number of photons $N=3\sinh^2r_1 + 1 \approx 300$. Note that such a highly-nonclassical  multi-photon quantum states could be very interesting for the non-Gaussian quantum optomechanics due to its much stronger interaction with the mechanical objects in comparison with \eg just single-photon ones.

The initial Wigner function is shown in Fig.~\ref{fig:WignerSinglePhot}\,a. It has a narrow negative area along the direction of squeezing (enclosed by the ellipse), therefore being even more susceptible to losses than a non-squeezed state. The conventionally reconstructed (without pre-amplification) Wigner function is plotted in Fig.\,\ref{fig:WignerSinglePhot}\,b, for the case of $1-\eta_d=0.05$. Here the shape of the Wigner function is highly distorted and only a very shallow negative-valued area remains. The reconstructed Wigner function after phase-sensitive amplification and loss is
\begin{equation}
 W^\prime_{\rm SSP}(q,p) = \frac{
      2q^2\dfrac{s\epsilon_r+1}{\epsilon_r+s} + 2p^2\dfrac{\epsilon_r+s}{s\epsilon_r+1}
      + \epsilon_r^2 - 1
    }{\pi\bigl[(\epsilon_r+s)(\epsilon_r+s^{-1})\bigr]^{3/2}}\,
    \exp\left(-\frac{q^2}{\epsilon_r+s}-\frac{p^2}{\epsilon_r+s^{-1}}\right)  ,
\end{equation}
which is plotted for the value of $r_2=2.3$ (20\,db), see Fig.\,\ref{fig:WignerSinglePhot}\,c. 

As another example, we consider the case of low detection efficiency $\eta_d=0.45$, which corresponds to $\epsilon=1.2$. In this case, the losses completely wash out the negative-valued area, see Fig.\,\ref{fig:WignerSinglePhot06}\,a. However, the pre-amplification allows one to recover the negativity, see Fig.\,\ref{fig:WignerSinglePhot06}\,b, and sufficiently strong pre-amplification restores the initial Wigner function, see Fig.\,\ref{fig:WignerSinglePhot06}\,c.

\begin{figure}[t]
  \includegraphics[width=0.62\textwidth]{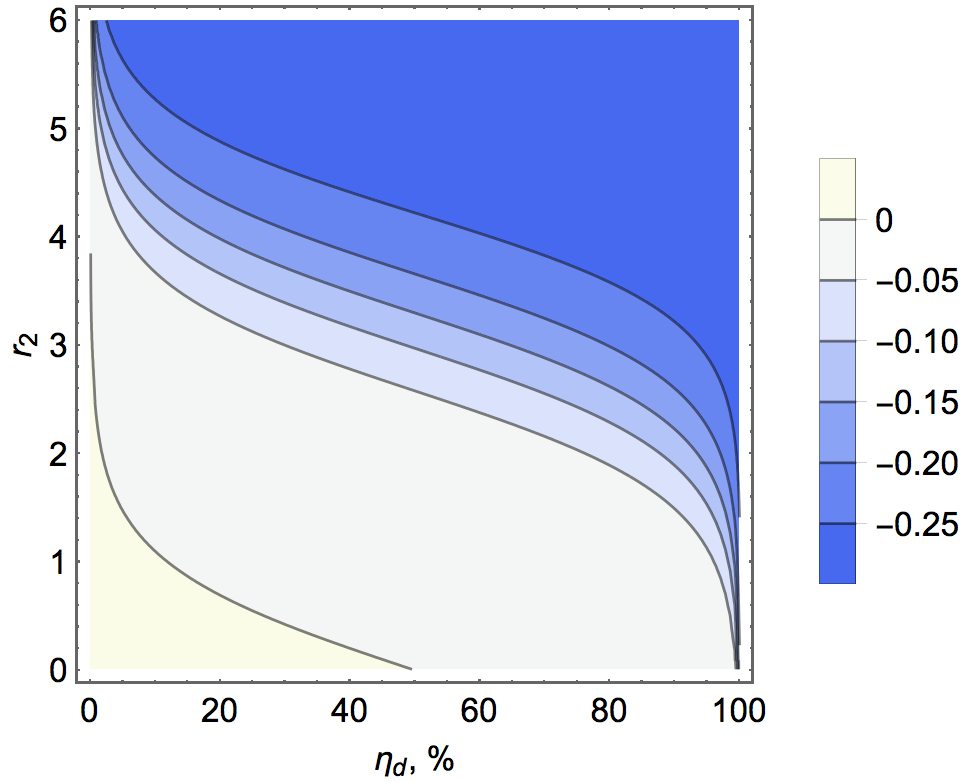}
  \caption{Depth of the Wigner function (\ref{eq:Depth}) for a squeezed single-photon state with $r_1=3$ (26 db of squeezing) as a function of the detector efficiency $\eta_d$ (given in \%) and the pre-amplification factor $r_2$. The plot is produced for the case of an imperfect BBO non-linear crystal.  Note that for any $\eta$, the ideal value of the depth equal to $-1/\pi$ could be reached by sufficiently strong perfect pre-amplification.}
\label{fig:Depth}
\end{figure}

As a quantitative measure of non-classicality for the squeezed single photon state, we consider the maximal negative depth of the Wigner function, which corresponds to its value at $q=p=0$ (see \eg \cite{Miwa_PRL_113_013601_2014}):
\begin{equation}\label{eq:Depth}
 W(0,0) = \frac{\epsilon_r^2-1}
    {\pi\left[(\epsilon_r+e^{2r_1})(\epsilon_r+e^{-2r_1})\right]^{3/2}} \, ,
\end{equation}
which we plot in Fig.\,\ref{fig:Depth} for the SSP state with $r_1=3$, as a function of the detector inefficiency $\eta_d$ and the pre-amplification factor $r_2$. Note that for squeezed single-photon state the existence of the non-classical negative-valued area does not depend on the initial squeezing $r_1$, but in the realistic lossy case its depth value decreases sharply with the increase of $r_1$: $|W(0,0)| \propto e^{-3r_1}$. Here again the perfect, or close to perfect pre-amplification allows, in principle, to restore the ideal value of the depth $-1/\pi$.

\section{Conclusion}

We have studied the protocol for the enhancement of the Wigner function tomography with real-world balanced homodyne detectors. The protocol relies on the phase-sensitive amplification of the quadrature under measurement before its homodyne detection. Our consideration includes the effect of losses in the nonlinear crystal serving as the traveling-wave phase-sensitive amplifier. We show that with this pre-amplification being sufficiently strong, one can reconstruct the quantum state close to the input one for any reasonable value of the detection loss. In particular, this protocol enables the observation of the Wigner-function negativity for a single-photon state under less than $50\%$ detector efficiency. As  practical examples, we considered bright squeezed vacuum and squeezed single-photon states, which are both strongly affected by optical losses and limited quantum efficiency. We showed that this protocol allows one to reconstruct the initial Wigner functions of these quantum states even in the presence of strong losses. This method promises considerable progress in future quantum optical and optomechanical experiments, especially with non-classical states.

\section*{Acknowledgments}
E. K. would like to thank Mathieu Manceau for key comments. This work was supported by the joint DFG-RFBR project CH1591/2-1 --- 16-52-12031 NNIOa. E. K. and F. K. acknowledge the financial support of the RFBR grant 16-52-10069.

\appendix

\section{ Absorption in the bulk of the non-linear crystal}\label{app:singlepass}

We treat the crystal as the set of $N\to\infty$ layers with the thickness $d/N\to0$, where $d$ is the total thickness of the crystal. The input/output relation for the amplified quadrature for layer $j$ is
\begin{equation}\label{layer}
(\hat{q}_\theta)_j  = \sqrt{1-\frac{kd}{N}}\,  (\hat{q}_\theta)_{j-1}  e^{r^\prime}
  + \sqrt{\frac{kd}{N}}\,    (\hat{q}_a)_j    \,,
\end{equation}
where $j=0\dots N$, $r^\prime=r_{\rm raw}/N$ is the parametric gain per layer in the absence of losses, $ (\hat{q}_\theta)_j $ is the amplified quadrature at the output of the $j$-th layer, and $ (\hat{q}_a)_j $ is the corresponding quadrature of the vacuum field injected into this layer due to the loss. Taking into account that the absorption per layer $kd/N\to0$, this equation can be recast as
\begin{equation}
   (\hat{q}_\theta)_j =   (\hat{q}_\theta)_{j-1}  e^{r/N} + \sqrt{\frac{kd}{N}}\,  (\hat{q}_a)_j  \,,
\end{equation}
where
\begin{equation} \label{eq:Reff}
  r = r_{\rm raw} - \frac{kd}{2}
\end{equation}
is the total effective squeezing, which is being detected in the experiment.

Using Eq.\,\eqref{layer} iteratively for $N$ layers, we obtain
\begin{equation}
   (\hat{q}_\theta)_N 
   =  [ (\hat{q}_\theta)_0 + \hat{q}_a] e^r \,, 
\end{equation}
where
\begin{equation}
  \hat{q}_a = \sqrt{\frac{kd}{N}}\,\sum_{j=1}^N   (\hat{q}_a)_j e^{-j r/N}
\end{equation} 
is the total noise introduced by the bulk and translated to the input of the  DOPA device.

Variances of all quadratures $ (\hat{q}_a)_j$ are equal to $1/2$ (vacuum state). Therefore, variance of $ \hat{q}_a$ is equal to 
\begin{equation}
  \sigma_a^2 = \frac{kd}{2N}\,\sum_{j=1}^N e^{-2j r/N}
  = \frac{kd}{2N}\,\frac{1 - e^{-2  r}}{e^{2  r/N} - 1} \,,
\end{equation} 
which in the limiting case of $N\to \infty$ reduces to (\ref{appeq:BulkNoiseVar}).

\bibliographystyle{phaip}
\bibliography{mqm,ligo,misc_u,biblio_u,abbots_my}

\end{document}